\pdfoutput=1
\documentclass[reprint,amsmath,amssymb,aps,prb,showpacs]{revtex4-1}

\usepackage{graphicx}
\usepackage[]{hyperref}
\hypersetup{
  colorlinks,
  linkcolor={blue},
  citecolor={blue},
  urlcolor={blue}
}

\begin{document}

\title{Aging in the three-dimensional Random Field Ising Model}
\author{Sebastian \surname{von Ohr}}
 \email{sebastian.von.ohr@uni-oldenburg.de}
\author{Markus Manssen}
\author{Alexander K. Hartmann}
\affiliation{Institute of Physics, Carl von Ossietzky University, 26111 Oldenburg, Germany}
\date{\today}

\begin{abstract}
We studied the nonequilibrium aging behavior of the Random Field Ising Model in three dimensions for various values of the disorder strength. This allowed us to investigate how the aging behavior changes across the ferromagnetic-paramagnetic phase transition. We investigated a large system size of $N=256^3$ spins and up to $10^8$ Monte Carlo sweeps. To reach these necessary long simulation times we employed an implementation running on Intel Xeon Phi coprocessors, reaching single spin flip times as short as 6\,ps. We measured typical correlation functions in space and time to extract a growing length scale and corresponding exponents.
\end{abstract}

\pacs{75.50.Lk, 75.40.Mg, 75.10.Hk}

\maketitle

\section{Introduction}
The Random Field Ising Model \cite{imry1975} (RFIM) has been studied extensively in theory \cite{nattermann1997} and in experiments \cite{belanger1997}, realized using dilute antiferromagnets. While the equilibrium properties are reasonably well understood, the dynamics is still under discussion. Of particular interest is the aging behavior, i.e., the dynamics resulting from starting in an equilibrium state followed by a rapid parameter change to a target point in phase space resulting in an non-equilibrium situation \cite{bouchaud1998,corberi2011}. Here we study the quench from a random configuration at infinite temperature to a low temperature with suitably small random-field disorder, which results in domains of parallel spins forming and growing over time. The disorder introduced by the random field pins the domain wall and slows down the growth of domains.

Typically, literature on aging concentrates on isolated target points in phase diagrams. Here we are interested in correlating the aging behavior with a disorder-driven phase transition. So far, aging in the RFIM has been analyzed mostly in the ferromagnetic phase \cite{rao1993, oguz1994, aron2008, corberi2012}. Some recent articles \cite{sinha2013, mandal2014} cover the two-dimensional (2D) RFIM in the disordered phase, but no studies of the 3D RFIM in a larger space of the phase diagram are known to the authors. Similar to previous work \cite{sgferro_ageing2015} analyzing the aging across the spin glass-ferromagnet transition in the Edwards-Anderson model, we here present results for the dynamic behavior for the 3D RFIM across the disorder-driven ferromagnet-paramagnet transition. We will be looking at the spatial correlation and discuss different methods of extracting the coherence length and corresponding exponents from it. We will also look at the autocorrelation and try to collapse it by rescaling the time in units of the coherence length, also yielding suitably defined exponents.

To reach sufficient long simulation times we implemented the model on the Intel Xeon Phi coprocessor. These cards offer performance comparable to Graphic Processing Units (GPUs), but the architecture is more similar to current CPUs, just with more cores. Parallelization does not require learning a new programming extension as for GPUs, but can be done using well-known techniques, e.g., OpenMP \cite{openmp}, MPI \cite{mpi} or just creating threads manually. Porting a simulation from CPU to Xeon Phi cards is straightforward; however, to fully utilize the performance a lot of knowledge about the architecture and careful optimization is necessary. Using our optimized implementation of the model we were able to simulate $10^8$ sweeps for 64 disorder samples of a large $N = 256^3$ system at many different values of the disorder strength.

The remainder of this article is structured as follows. In Sec.~\ref{sec:model} we describe the RFIM and the observables used to characterize the aging of the system. Section~\ref{sec:implementation} describes details of the implementation on Xeon Phi cards. Results of the simulation are presented in Sec.~\ref{sec:results}. We close with our conclusions in Sec.~\ref{sec:conclusion}.

\section{Model} \label{sec:model}
\begin{figure}
 \includegraphics{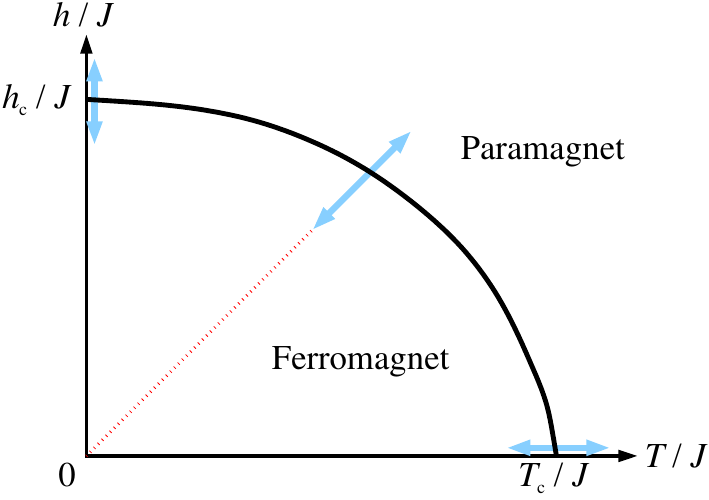}
 \caption{(Color online) Sketch of the phase diagram: for low temperatures and low disorder, the system is in a ferromagnetic phase, otherwise in a paramagnetic phase. \label{fig:phase}}
\end{figure}

The Random Field Ising Model describes a $D$-dimensional cubic system of side length $L$ containing $N = L^D$ Ising spins $S_i = \pm1$. The Hamiltonian is given by
\begin{equation}
 H(S) = -J \sum_{\langle i, j \rangle} S_i S_j - \sum_i h_i S_i
\end{equation}
where the sum runs over nearest neighbors $\langle i, j \rangle$ and the field $h_i = h_0 \varepsilon_i$ with $\varepsilon_i = \pm 1$ being a quenched random variable. Here, we apply a symmetric bimodal distribution $P(\varepsilon_i) = [\delta(\varepsilon_i - 1) + \delta(\varepsilon_i + 1)]/2$. The boundary conditions are periodic in all directions. The parameter $h_0$ controls the strength of the random field. For $h_0 = 0$ the well known pure Ising Model is reproduced with a paramagnetic phase at high temperatures $T > T_\text{c}$ and a ferromagnetic phase at low temperatures. With $h_0 > 0$ additional disorder is introduced, which lowers the transition temperature to the paramagnetic phase. In Fig.\ \ref{fig:phase} a phase diagram for $D \ge 3$ dimensions is shown \cite{universality2013}. Even at $T = 0$ the system is in a paramagnetic phase if the field strength becomes too large $h_0 > h_\text{c}$. In the remainder of the article will only be concerned with the case $D = 3$, which has a zero-disordered critical temperature $T_\text{c} \approx 4.5115$ \cite{binder2001} and a zero-temperature critical field strength $h_\text{c} \approx 2.20$ \cite{art_uli1999, malakis2008}.

Simulations start with random initial configurations, emulating a quench from infinite temperature. A single-spin flip Monte Carlo simulation is performed, see below for details. We then examine the system at different waiting times $t_\text{w}$ (measured in Monte Carlo sweeps) after the beginning of the simulation. The order parameter is the magnetization
\begin{equation}
 m = \frac{1}{N} \sum_i S_i \,.
\end{equation}
To measure the growing length scale we make use of the spatial two-point correlation
\begin{equation}
 C_2(r, t_\text{w}) = \frac{1}{N} \sum_i S_i(t_\text{w}) S_{i+r}(t_\text{w})
\end{equation}
between two points. With $i+r$ we denote a spin, which has a spatial distance $r$ from spin $i$.

There exist different approaches to extract a growing coherence (or dynamic correlation) length $\xi$ from the spatial correlation function, for a recent comparison in the case of the three-dimensional random-bond (spin-glass) model see, e.g., Ref.~\onlinecite{sgferro_ageing2015}. Most approaches are based on the assumption that $C_2$ follows the functional form
\begin{equation} \label{eqn:spatialC}
 C_2(r, t_\text{w}) \propto r^{-\alpha} g\!\left( \frac{r}{\xi(t_\text{w})} \right) \,,
\end{equation}
where the function $g$ is approximately a stretched exponential $g(x) \approx \exp(-x^\beta)$. Extraction of $\xi$ and corresponding exponents works by fitting \eqref{eqn:spatialC} to the data of $C_2$.

Alternatively one can use integral estimators, first used for spin-glasses \cite{belletti:2008}, which work without assuming the functional form of $g$. This is done by calculating the integral
\begin{equation} \label{eqn:Ik}
 I_k(t_\text{w}) = \int_0^{L/2} r^k C_2(r, t_\text{w}) \,\text dr
\end{equation}
which allows the calculation of the coherence length using
\begin{equation} \label{eqn:intEst}
 \xi_{k,k+1}(t_\text{w}) = \frac{I_{k+1}(t_\text{w})}{I_k(t_\text{w})} \propto \xi(t_\text{w}) \,.
\end{equation}
A value of $k = 1$ is recommended in Ref.~\onlinecite{belletti:2008} as a tradeoff between systematic errors for low $k$ values and statistical errors for larger $k$. This method also allows to determine the exponent $\alpha$ since $I_1 \propto \xi_{1,2}^{2 - \alpha}$.

A different, but very simple, method of extracting the coherence length uses the inverse density of defects \cite{corberi2012}. A defect is a spin with at least one antiparallel neighbor. With the number of defects $D(t_\text{w})$ the coherence length is given by $\xi(t_\text{w}) = N/D(t_\text{w})$.

Another observable of interest is the autocorrelation
\begin{equation} \label{eqn:autocor}
 C(t, t_\text{w}) = \frac{1}{N} \sum_i S_i(t_\text{w}) S_i(t_\text{w} + t)
\end{equation}
comparing the same system at different times. It is expected to split into two parts. The first quasi-equilibrated part for $t \ll t_\text{w}$ takes the form \cite{rieger1996,kisker1996,komori2000,berthier2002,jaubert2007} of a power law
\begin{equation} \label{eqn:auto_eq}
 C_\text{eq}(t) \propto t^{-x}
\end{equation}
with a characteristic exponent $x$. The later aging part $C_\text{age}(t, t_\text{w}) = f( \xi(t_\text{w} + t) / \xi(t_\text{w}) )$ is expected \cite{bouchaud1998,corberi2011} to depend only of the ratio of the coherence lengths at the two times. For long waiting times $\lim_{t_\text{w} \to \infty}C(t, t_\text{w}) = C_\text{eq}(t) + B^2$ a plateau is expected \cite{huse1987,golubovic1991}, with $B$ equal to the equilibrium magnetization $M$ in the ferromagnetic phase.

\section{Implementation} \label{sec:implementation}
We implemented a standard Metropolis Monte Carlo simulation \cite{newman} of the model for the Intel Xeon Phi 3120P coprocessor. It is based on an earlier implementation for GPUs, but in the remainder of the article we cover only the implementation for Xeon Phi cards. For details of a similar GPU implementation see Ref.~\onlinecite{sgferro_ageing2015}. As a reference for the Xeon Phi architecture we refer to the official documentation \cite{intel:system_software, intel:vpu}. It's main points are the 57 Pentium based cores with 4-way Hyper-Threading and the 512\,bit wide vector processing unit.

\begin{figure}
 \includegraphics{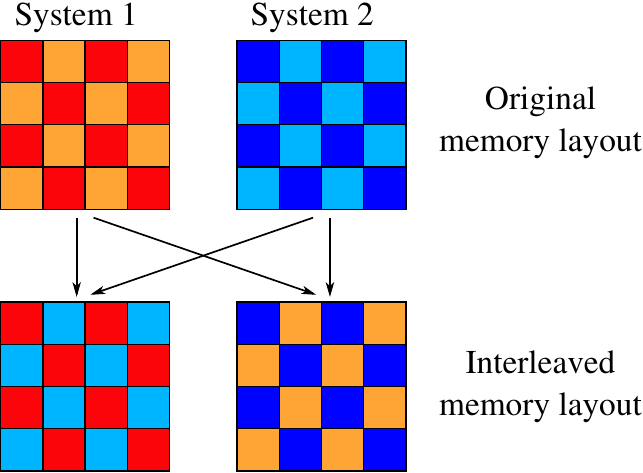}
 \caption{(Color online) Example memory layout of a 2D system, before and after exchanging the odd (light color) spins. Each spin in one interleaved memory block can be updated independently because all neighbors are stored in the other memory block. \label{fig:spin_exchange}}
\end{figure}

In each sweep every spin is updated as follows. First, the energy change
\begin{equation}
 \Delta E_i = 2 J S_i \sum_{j \in N(i)} S_j + 2 h_i S_i
\end{equation}
for a flip of spin $i$ is calculated. The sum runs over the neighboring sites $N(i)$ of spin $i$. Next, a random number is generated and the spin $i$ is flipped with probability
\begin{equation} \label{eqn:flip_prob}
 p_\text{accept} = \min\!\left[1, \exp\!\left( {-\Delta E_i}/{T} \right) \right] \,.
\end{equation}
To keep all cores of the Xeon Phi card busy, spins need to be updated simultaneously. If each thread would update a cubic subset of the system in a linear fashion then spins on the border of each subset have neighbors in a different subset and thus depend on data owned by a different thread. If threads are not synchronized these neighboring spins may change in a non-deterministic way, generating irreproducible results. Instead of synchronizing threads on the border, we chose to update spins in a checkerboard pattern. First, all even (sum of the $x, y, z$ coordinates are even) spins are updated and the odd spins stay the same. Next, threads are synchronized and the roles are switched and all odd spins are updated. However, this update scheme results in an unfavorable access pattern to the memory, since we want to use SIMD instructions and load large chunks of continuous data at a time. Instead, we simulate two systems simultaneously and interleave them in memory, so that the even spins from the first system are in same memory block as the odd spins from the second system. Fig.~\ref{fig:spin_exchange} shows how spins would be exchanged in a 2D system.
With this technique one can update all spins in one memory block because all the neighboring spins are stored in the other block. This also simplifies the spin update routine because spins can be updated in a linear fashion again, which effectively alternates between two samples. One sweep consists of updating both memory blocks.

We also implemented multispin coding, which means, that each spin is represented by a single bit and stored together with other spins in a single byte or some bigger data type. We used a 32\,bit data type to store 32 spins of the same location but of different samples. Since the random field only takes the values $\pm h_0$ it can also be encoded using the same technique. Together with the memory interleaving this results in 64 samples for each run of the simulation. The calculation of the energy difference is then mapped to bitwise logic operations. In the case of the 3D RFIM there are a total of 14 possible energy differences (spin aligned with $6,\dotsc,0$ neighbors, i.e., seven cases, times two cases for alignment or antialignment with field) and the spin flip probability can be precalculated for these cases and stored in a lookup table. The only thing that cannot easily be mapped to bitwise logic operations is the random number generation. So, we just use the same random number for 32 samples.

Most traditional pseudorandom number generators (PRNG) like the Mersenne Twister \cite{matsumoto1998} are unsuitable for highly parallel architectures because each instance requires comparative large amounts of memory, i.e., about 2.5\,kB for the Mersenne Twister. Depending on the PRNG it's also nontrivial to initialize multiple instances of the same generator so that the output streams do not overlap for a sufficient number of calls. Instead, we used a counter-based random number generator \cite{salmon2011}, namely the Philox PRNG. The difference compared to traditional generators is that they do not operate one some internal state, which is advanced with every number generated. Instead, they consist of a deterministic function, taking a key and a counter as parameters, and return a random number generated from those parameters. The key is chosen so that it's different in each thread and the counter is incremented after each function call. In our simulation we used the index $0 \dots N-1$ of the spin as the key. This effectively reduces the state of the PRNG to a single counter, which is shared across all threads. Since the reference implementation by the authors of Ref.~\onlinecite{salmon2011} doesn't support the Xeon Phi architecture, we implemented our own optimized version using SIMD instructions.

With the described optimizations we reached an effective single spin-flip attempt time of ${\approx}6$\,ps on the Xeon Phi card, corresponding to 57 (processors) $\times 32$ (spins per integer) $\times 16$ (vector processing unit) parallel flip attempts per $1.8 \cdot 10^{-7}$\,s, equal to roughly 180 cycles at the $10^{-9}$\,s cycle time of the processors on the card. Compared to ${\approx}9$\,ps of our previous implementation on a GeForce GTX 570 GPU the Xeon Phi card implementation is faster. But this was expected since the GPU is older than the Xeon Phi 3120P card. Overall, the performance of the Xeon Phi card seems to be comparable to GPUs. One advantage of the Xeon Phi card was, however, that the reported spin flip time was archived with only 64 samples while the GPU implementation used 128 samples. This is because operations with 32\,bit integers on the GPU are only marginal faster than operations with 64\,bit integers, while on the Xeon Phi card 32\,bit calculations are twice as fast as 64\,bit operations. So, reducing the number of samples on the GPU to 64, increases the single spin flip time to ${\approx}13$\,ps.

\section{Results} \label{sec:results}
For system size $N=256^3$ we simulated 64 samples of randomly initialized realizations. Simulations were performed on 8 Intel Xeon Phi 3120P cards. The temperature was fixed to $T = 0.8$ and the field strength was chosen in the range $h_0 \in [1.2, 3.2]$, covering the ferromagnetic and paramagnetic phase. We do not know the precise critical field strength at $T = 0.8$, but analysis of the phase diagram \cite{universality2013} suggest that the $T=0$ value $h_\text{c}$ is still very close to the real critical field strength at $T = 0.8$. In most simulations $10^8$ sweeps were performed, which took about a week on a single Xeon Phi card. At specific waiting times $t_\text{w}$ we saved the whole spin configuration to disk, so that they can be post-processed to calculate the observables. For $h_0 \ge 2.4$ we performed only $10^7$ sweeps, mainly because they are close to equilibration and would require a large amount of disk space since the disordered spin configurations can not be compressed very well.

\begin{figure}
 \includegraphics{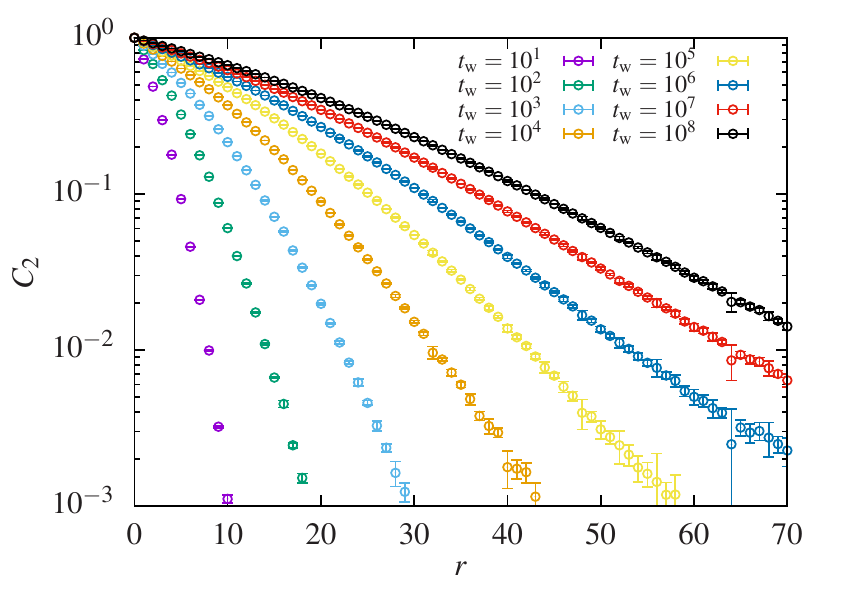}
 \caption{(Color online) Spatial correlation $C_2$ over the distance $r$ for different waiting times $t_\text{w}$ of a $256^3$ system with field strength set to $h_0 = 1.6$. Only data points at integer values of $r$ are shown to give a clearer picture. \label{fig:splot}}
\end{figure}

An exemplary spatial correction $C_2$ for $h_0 = 1.6$ is shown in Fig.~\ref{fig:splot}.
The curves show an almost exponential decay with a slight bend, which is captured by the exponent $\beta$. The spatial correlation decreases more slowly for larger waiting times, suggesting a growing length scale.

\begin{figure}
 \includegraphics{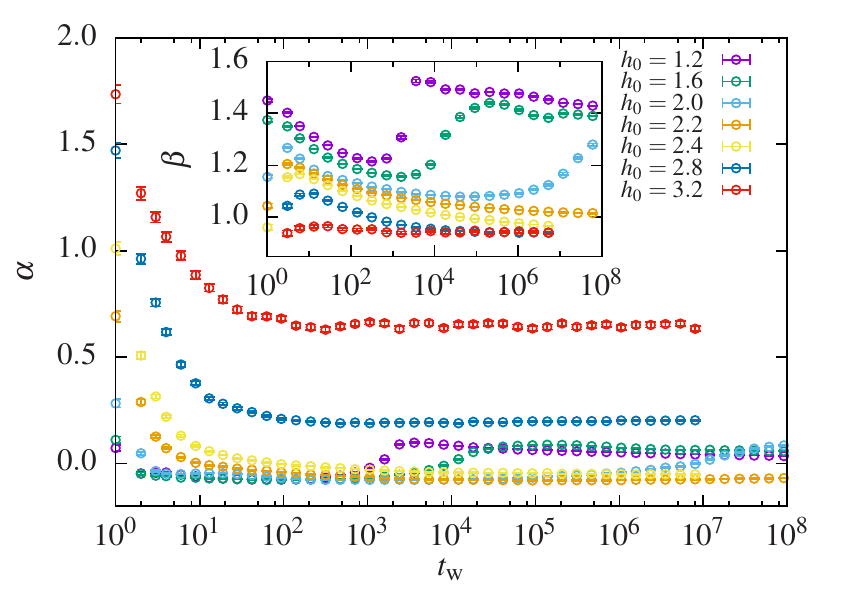}
 \caption{(Color online) Exponent $\alpha$ and $\beta$ (inset) extracted using individual fits to the spatial correlation $C_2$ at waiting time $t_\text{w}$. The range of the fit was set to $r \in [1, 64]$. \label{fig:alpha_beta_comb}}
\end{figure}

\begin{figure}
 \includegraphics{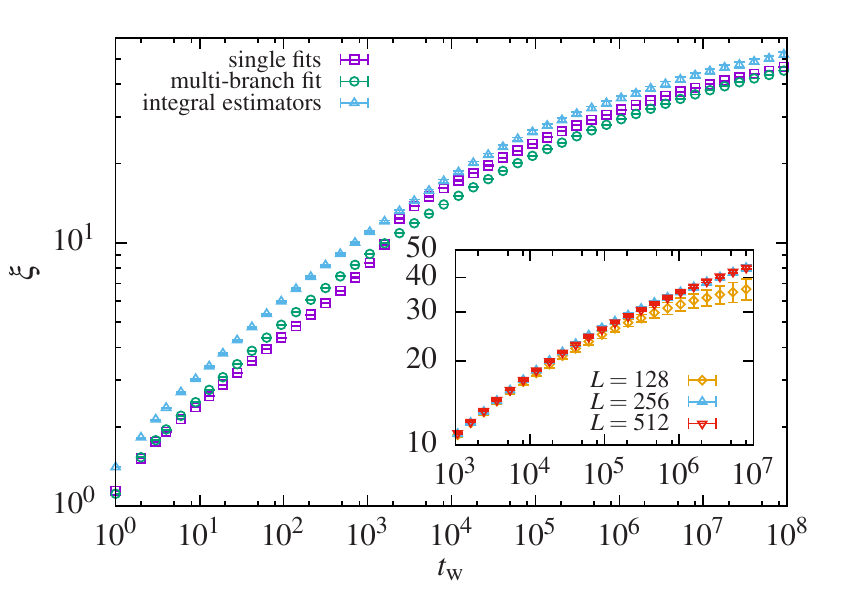}
 \caption{(Color online) Comparison of coherence length $\xi$ extracted using different methods at $h_0 = 1.2$. Inset: Growth of coherence length extracted using the integral estimators for different system sizes $L$. \label{fig:comparison}}
\end{figure}

Extracting the coherence length from the data proved to be complicated and different approaches were tried. At first a fit with \eqref{eqn:spatialC} was tested. The exponent $\alpha$ is expected to be $0$ for the RFIM, but was added as a parameter nevertheless since previous works only analyzed the system in the ferromagnetic region. Since the fit function diverges at $r = 0$ for an exponent $\alpha > 0$ the fit has to be restricted to a smaller range $r \ge r_\text{min} > 0$. The choice of $r_\text{min}$ strongly affects the result of the fit and has to be chosen carefully, as we discuss next extensively. First, Fig.~\ref{fig:alpha_beta_comb} shows the resulting value of $\alpha$ and $\beta$ by using single fits to each measured spatial correlation at waiting time $t_\text{w}$.
We used a small value of $r_\text{min} = 1$, to get better results for short times and high disorder values, since only few data is available here. In Fig.~\ref{fig:comparison} the resulting values of the coherence length are shown. For disorder values $h_0 \le 2.0$ the exponent $\beta$ shows a large jump at later times. The nature and position of the jump is dependent on details of the fitting procedure. If a bigger $r_\text{min} = 3$ is chosen (not shown) the curves for $h_0 \le 2.0$ start at a much higher value $\beta \approx 2$ and then decrease until they join with the $r_\text{min} = 1$ curves around the position of the jump. Also, by ignoring the error bars of the spatial correlation for the fit, the jump shifts to earlier times. All other curves for $\beta$ show a monotonic decrease after an initial growth below $t_\mathrm{w} \le 10$. Although the actual values resulting from the fit at smaller waiting times depend on the details of the fitting procedure, a general picture is visible: The results show a consistent behavior in both phases. In the ferromagnetic phase the curves tend with increasing waiting time towards $\beta \approx 1.4$ while in the paramagnetic phase the exponent $\beta$ seems to be $1$.

The exponent $\alpha$ as a function of the waiting time $t_\text{w}$ shows a similar jump, but less pronounced. The fitted $\alpha$ is also negative for some waiting times, but constraining $\alpha$ to positive values does not give a better fit quality. The exponent is close to 0 in the ferromagnetic phase, just as expected, but increases in the paramagnetic phase. Overall, we conclude that the single fit method is not a reliable method to extract exponents, at least for rather short waiting times, because it depends strongly on the fit range and other details.

Since we found that the individual fits using different suitably chosen values of $r_\text{min}$ agree on the exponents for large waiting times, where the exponents are mostly constant, we may assume that they are constant for all waiting times. Therefore, we tried using a multi-branch fit \cite{practical_guide2015}, which uses common exponents for all waiting times. By fitting to all spatial correlations simultaneously, the hard to extract exponents of the early waiting times are influenced by later waiting times with more data. This also prevents sudden (slight) changes in the fit values of the coherence length in \eqref{eqn:spatialC}, as visible in Fig.~\ref{fig:comparison}, because it's no longer influenced by changing exponents. If the assumption, that the exponents for a given $h_0$ are constant, is correct then this is a much better method to extract the coherence length, than using individual fits.

The last approach uses \eqref{eqn:intEst} to calculate $\xi_{1,2}$. Integrals $I_k$ are calculated by numerically integrating the data until the value first becomes smaller than three times it's error. The remaining part of the integral is approximated by fitting \eqref{eqn:spatialC} to the data and then integrating the fitted function. The choice of $r_\text{min}$ is less critical here since only the tail of the fitted function is used and the contribution to the integral is small. Extracting the coherence length using this method is straightforward and yields very similar results compared to the multi-branch fit. Only for very short waiting times there a minor differences. Also, the coherence length extracted using the integral estimators is only proportional to the one extracted using the multi-branch fit. The factor depends on the disorder value since the multi-branch fit seems to be more susceptible to the exact shape of the spatial correlation. Fig.~\ref{fig:comparison} shows a comparison of the different methods to extract the coherence length.
The curves from the multi-branch fit and the integral estimator are basically the same, except for a constant shift, which is because of the unknown prefactor in the integral method. Thus, we can safely use from now on the coherence length obtained from the integral method. Note that the curve from the single fits shows a jump around $t_\text{w} = 10^3$ which is due to the difficulty to fit to few data points and the above discussed behavior of the exponents $\alpha$ and $\beta$.

The inset of Fig.~\ref{fig:comparison} shows a comparison of the coherence length for different system sizes. For up to $t_\text{w} = 10^7$ we simulated a $L = 512$ system and the extracted coherence length is essentially the same as for the $L = 256$ system. Therefore, we believe that the observables are free from finite size effects, even for $t_\text{w} = 10^8$. For a smaller $L = 128$ system there is a clear deviation from the other two curves.

\begin{figure}
 \includegraphics{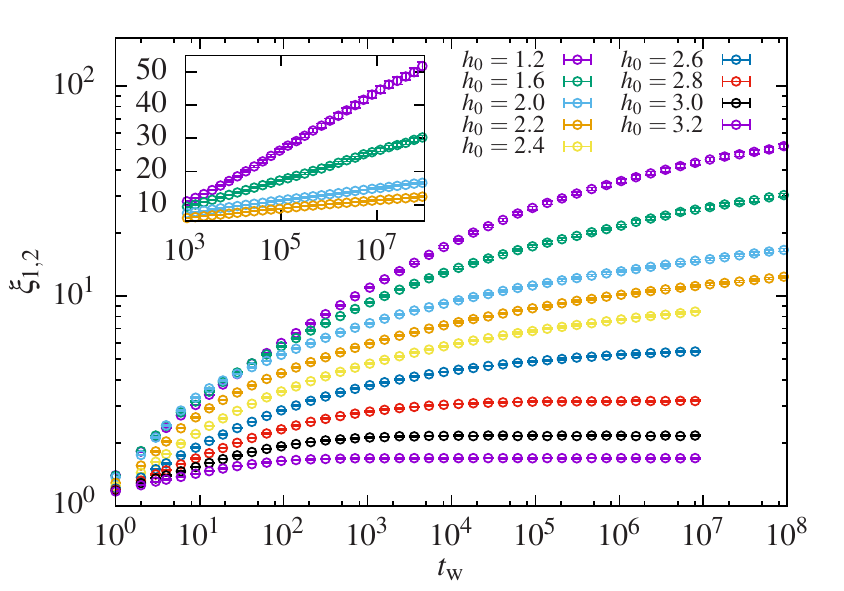}
 \caption{(Color online) Coherence length $\xi_{1,2}$ over the waiting time $t_\text{w}$ for different field strengths $h_0$. The coherence length was extracted from the spatial correlation using integral estimators. Inset: Section of the same data on a semi-log scale.\label{fig:dplot}}
\end{figure}

The extracted coherence length is shown in Fig.~\ref{fig:dplot} for different disorder values.
Because comparatively high field strengths $h_0$ were chosen, the curves do not show the usual power law behavior, but instead show a possibly logarithmic growth. The curves for $h_0 \ge 2.8$ equilibrate within the simulated waiting time. In the beginning the curves for $h_0 = 1.6$ and $h_0 = 2.0$ seem to grow faster than the $h_0 = 1.2$ curve, but this is a result of the undetermined prefactor.

\begin{figure}
 \includegraphics{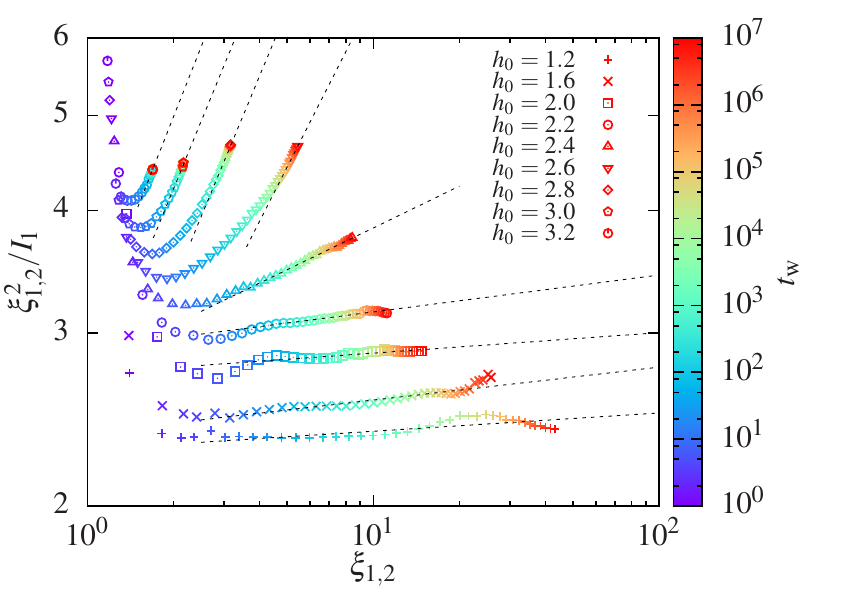}
 \caption{(Color online) Power law fits (dashed lines) to extract the exponent $\alpha$ from the integral estimators. The color gradient shows from which waiting time the values of $\xi_{1,2}$ and $I_1$ were extracted. \label{fig:alpha_pow}}
\end{figure}

\begin{figure}
 \includegraphics{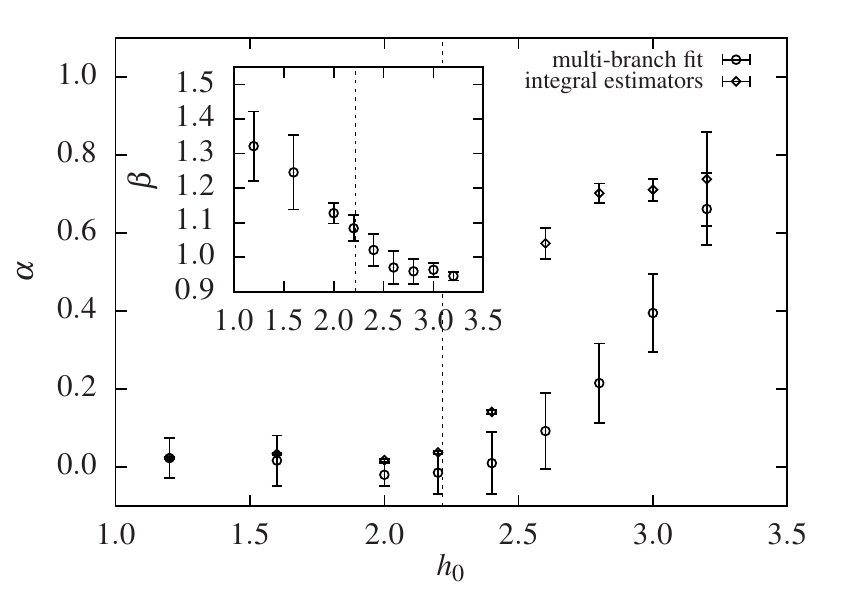}
 \caption{Scaling exponent $\alpha$ of the spatial correlation over the field strength $h_0$ for a $256^3$ system, extracted using different methods. The vertical line marks $h_0 = h_\text{c}$. Inset: Associated scaling exponent $\beta$. \label{fig:alphabeta}}
\end{figure}

The integral estimators also provides a means to extract the exponent $\alpha$, using the relation $I_1 \propto \xi_{1,2}^{2 - \alpha}$. By plotting $\xi_{1,2}^2 / I_1$ over $\xi_{1,2}$, as depicted in Fig.~\ref{fig:alpha_pow}, the exponent can be extracted using a power law fit.
It is observed, that the curves do not show a clear power law, especially for small $\xi_{1,2}$. However, for larger $\xi_{1,2}$ all curves become more or less straight lines. Power laws were fitted to those parts, subjectively choosing the beginning of the fit range for every curve. It can be seen that there is a clear crossover from a slope close to zero for $h_0 \le 2.2$ to a larger slope for larger $h_0$. Note that for high disorder values the system equilibrates and the coherence length stops growing. For these curves only the short waiting times contribute to the power law, as it can be seen from the color gradient. The resulting exponent $\alpha$ for different disorder values is displayed in Fig.~\ref{fig:alphabeta}, together with the exponents $\alpha$ and $\beta$ extracted using the multi-branch fit.
The different methods to extract $\alpha$ mostly agree, except for the range $h_0 \in [2.4, 3]$ where the $\alpha$ from the integral estimators grows faster. Below the critical field strength the exponent $\alpha$ is zero within the error margin, just like expected. In the paramagnetic region $\alpha$ grows quickly with increasing $h_0$. We also tried to extract the exponent $\beta$ again using a fit with \eqref{eqn:spatialC}, but with the parameters $\alpha$ and $\xi$ fixed to the already extracted values from the integral estimator and with an additional adjustable prefactor. Because of the discrepancies for the exponent $\alpha$, no reasonable fit was possible in the $h_0 \in [2.4, 3]$ range. Therefore, we only displayed the $\beta$ results from the multi-branch fit. The exponent $\beta$ also undergoes a change near $h \approx 2.6$, i.e., a bit beyond the phase transition.

\begin{figure}
 \includegraphics{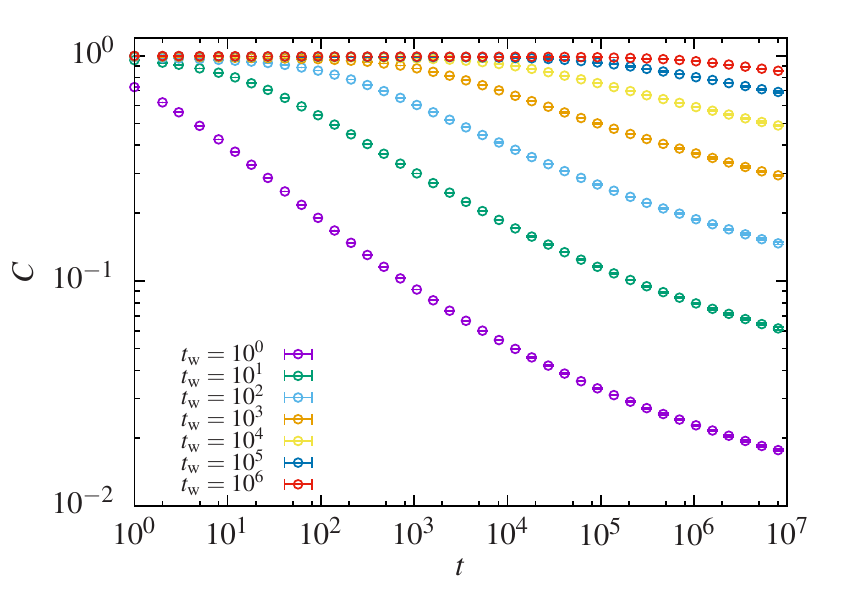}
 \caption{(Color online) Autocorrelation $C$ as a function of time $t$ for different waiting times $t_\text{w}$ at field strength $h_0 = 1.2$. \label{fig:autocor_1_2}}
\end{figure}
\begin{figure}
 \includegraphics{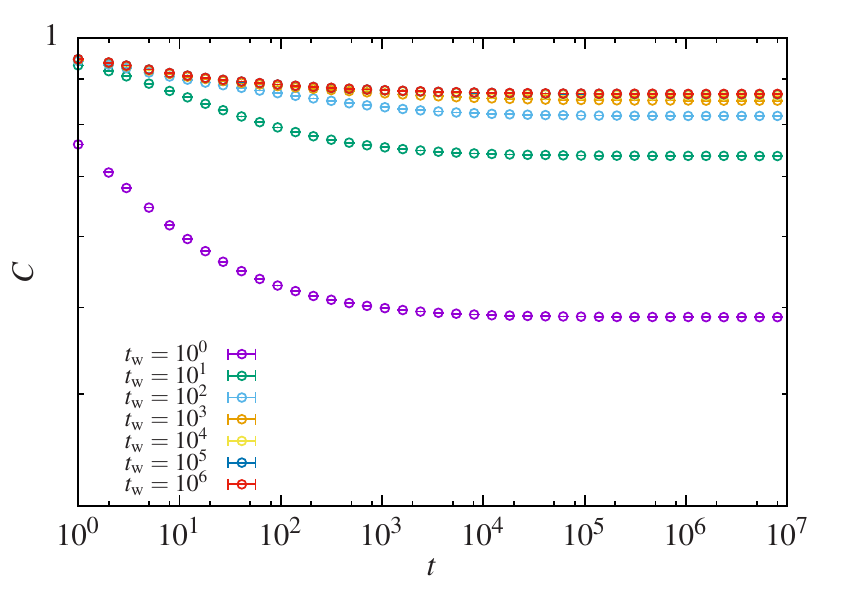}
 \caption{(Color online) Autocorrelation $C$ as a function of time $t$ for different waiting times $t_\text{w}$ at field strength $h_0 = 2.8$. Note that the data for $t_\text{w}=10^4$, $10^5$ and $10^6$ falls on top of each other. \label{fig:autocor_2_8}}
\end{figure}
\begin{figure}
 \includegraphics{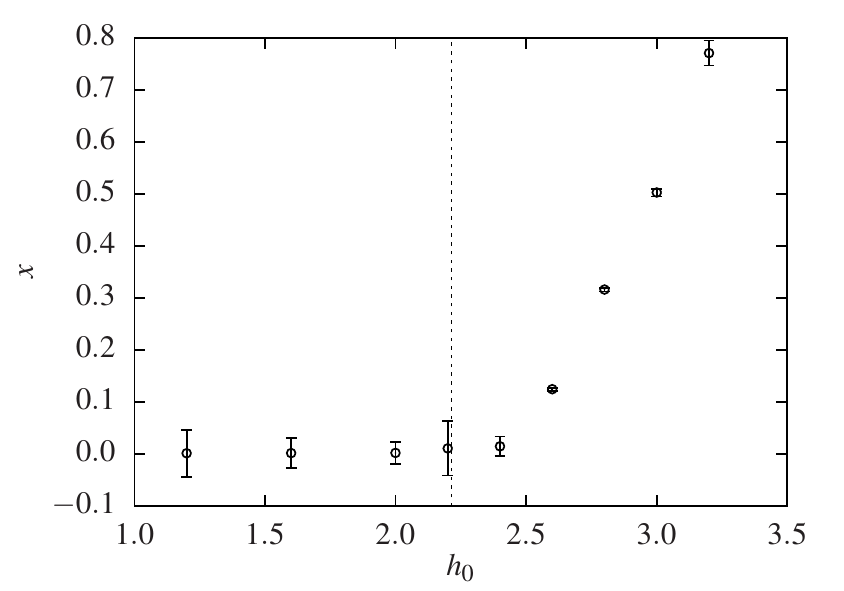}
 \caption{Equilibrium exponent $x$ for different values of disorder $h_0$. The vertical line marks $h_0 = h_\text{c}$. \label{fig:exp_x}}
\end{figure}

Next, we look at the autocorrelation from \eqref{eqn:autocor}. An exemplary curve for $h_0 = 1.2$ is shown in Fig.~\ref{fig:autocor_1_2}. The two parts of the autocorrelation with a transition around $t \approx t_\text{w}$ can be observed. The first quasi-equilibrated part looks like a constant function while the second aging part first decays as a power law which then slows down for short waiting times. In the paramagnetic region the system equilibrates and the autocorrelation relaxes to a plateau, as can be seen in Fig.~\ref{fig:autocor_2_8} for the case $h_0=2.8$.
This is because the spins are mostly aligned in the direction of the field and the autocorrelation shows only the fluctuation around this configuration. To extract the exponent $x$ of the quasi-equilibrated part \eqref{eqn:auto_eq} a fit of the form $C_\infty(t) = A \cdot t^{-x} + B^2$ was performed. In the paramagnetic phase the value of $B$ can be easily read from the height of the plateau. The fit range was chosen by plotting $C(t) - B^2$ on a log-log scale and restricting the fit to the straight part of this curve. In the ferromagnetic phase the exponent $x$ is close to 0 and doesn't allow to determine $B$ with reasonable accuracy. Therefore, we restricted the fit to the straight part of $C(t, t_\text{w})$ and show only the results for the exponent $x$ in Fig.~\ref{fig:exp_x}.
Here again the change in dynamics is visible slightly above the critical field strength.

\begin{figure}
 \includegraphics{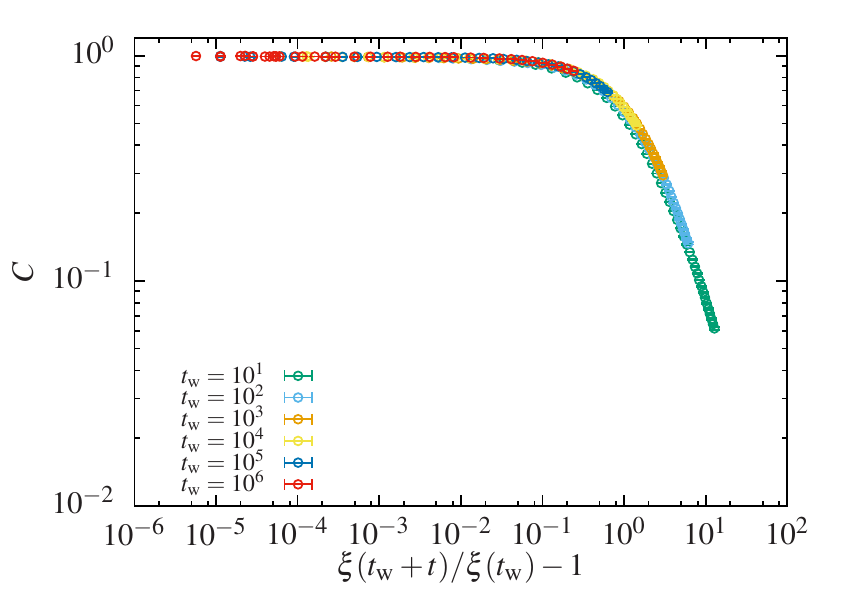}
 \caption{(Color online) Collapse of autocorrelation $C$ for different waiting times $t_\text{w}$ at field strength $h_0 = 1.2$. \label{fig:autocor_1_2_col}}
\end{figure}

Last we check the assumption that the aging part scales with $\xi(t_\text{w} + t) / \xi(t_\text{w})$. We used the coherence length $\xi_{1,2}$ extracted using the integral estimators. In the ferromagnetic phase we can just ignore the quasi-equilibrated part, because it is close to a constant function, and do a collapse of $C$ directly, as depicted in Fig.~\ref{fig:autocor_1_2_col} for the case $h_0=1.2$. We subtract one from the abscissa to make the collapse for values $t \ll t_\text{w}$ better visible. It can be seen that the quality of the collapse is very good.

\begin{figure}
 \includegraphics{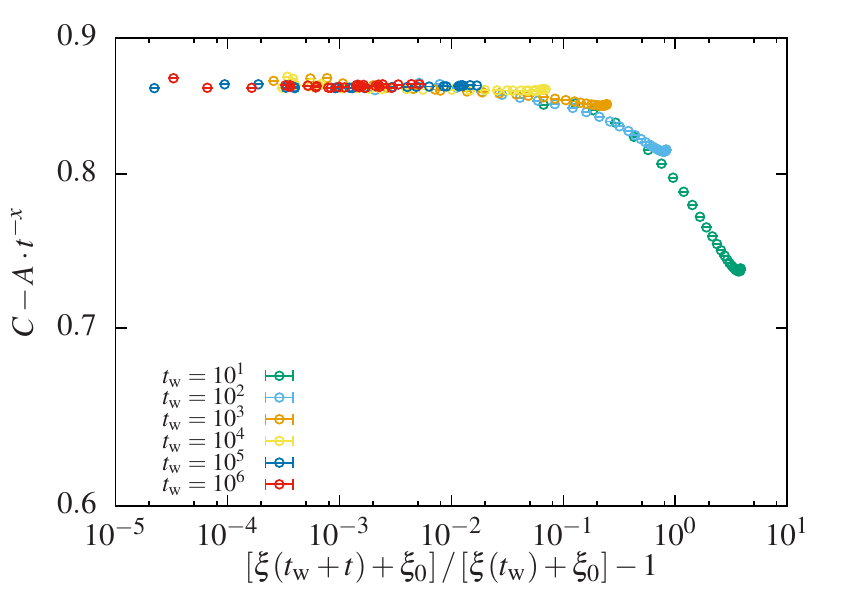}
 \caption{(Color online) Collapse of autocorrelation $C$ with correction for the quasi-equilibrated part $C_\text{eq}$ for different waiting times $t_\text{w}$ at field strength $h_0 = 2.8$. An additional additive constant $\xi_0 \approx -1.333$ was added to the coherence length to give a better collapse. \label{fig:autocor_2_8_col}}
\end{figure}

In the paramagnetic phase the quasi-equilibrated part is not constant and has to be accounted for in the collapse. We tested an additive decomposition $C(t, t_\text{w}) = C_\text{eq}(t) + C_\text{age}(t, t_\text{w})$ and also a multiplicative decomposition $C(t, t_\text{w}) = C_\text{eq}(t) \cdot C_\text{age}(t, t_\text{w})$. We found that the additive decomposition gave a better collapse in the $t \ll t_\text{w}$ range, as shown in Fig.~\ref{fig:autocor_2_8_col}.
The parameters $A$, $x$ and $\xi_0$ were adjusted for the best collapse. For this example the parameter $x$ is close to the one extracted using a fit, as shown in Fig.~\ref{fig:exp_x}, but in the ferromagnetic phase the collapse gives unrealistic large values for $x$. Except for the tails of the individual curves the collapse works reasonably well. The slight increase toward the end can probably be attributed to the system reaching equilibrium.

\section{Conclusion} \label{sec:conclusion}
For the Random Field Ising Magnet, we performed long-time Monte Carlo simulations for large systems ($N=256^3$) at low temperature and different disorder strengths $h_0$, covering the ferromagnetic and paramagnetic phase. This was made possible by the usage of several Intel Xeon Phi cards and an optimized implementation of the model. Getting the first working simulation on the Xeon Phi card was relatively easy because of the similar architecture to CPUs. However, such a naive implementation was slower than the same naive implementation on a current CPU. The main reason for this seems to be the poor usage of the vector processing unit. Therefore, we rewrote most of the code and directly accessed the vector processing unit by using SIMD intrinsics. Implementing the same model on GPUs was more complicated in the beginning, but already the first version outperformed the CPU. We observed that the overall time spent optimizing for the Xeon Phi card and GPUs is roughly the same.

By using this implementation, we were able to study the aging behavior of the model. We analyzed the results by looking at the spatial correlation and the autocorrelation. The coherence length was extracted from the spatial correlation using a number of different methods. Here the integral estimators introduced in Ref.~\onlinecite{belletti:2008} proved to be the most reliable approach. Also, a multi-branch fit, restricting the exponents of one run to the same values, was superior to individual fits. Moving into the paramagnetic phase complicated the analysis since the spatial correlation is very short and the system equilibrates within the simulated timespan. At the ferromagnet-paramagnet phase transition we found a clear change of the aging behavior, as visible in the exponents $\alpha$, $\beta$ and $x$. However, the exponent $\beta$ and $x$ show a transition slightly above $h_\text{c}$, which might indicate a connection to a percolation transition like that found slightly above the phase transition for the similar RFIM with Gaussian disorder \cite{seppala2002}. Note that in a previous study \cite{sgferro_ageing2015} of the Edwards-Anderson model a change in dynamics was found that coincides fully with the ferromagnet-spin glass transition, i.e., it was visible in all measured exponents simultaneously.

The autocorrelation splits into a quasi-equilibrated part, which follows a power law with the exponent $x$, and a later aging part. In the ferromagnetic phase $x$ is close to 0 and starts growing in the paramagnetic phase. Thus, we can mostly ignore the quasi-equilibrated part in the ferromagnetic phase and do a collapse of the aging part by plotting the autocorrelation over the quotient $\xi(t_\text{w} + t) / \xi(t_\text{w})$. In the paramagnetic phase the collapse is becoming more difficult with increasing disorder strength. We found that an additive decomposition of the autocorrelation and an additive constant to the coherence length gave the best collapse.

Overall, we found that the equilibrium phase-transition behavior of the model and maybe a particular percolation transition are well reflected in the dynamic observables. Due to the availability of relatively inexpensive yet powerful GPU and Intel-Phi architectures, one could easily extend our studies to the RFIM in higher dimensions \cite{rfim4d2002, mean-field2011}, to RFIM systems with correlations \cite{RFIM_correl2011}. Furthermore, it could be interesting to investigate whether the aging behavior can be understood in terms of non-trivial low lying excitations \cite{fes_rfim2008}. Certainly, it would be interesting to perform such studies for other spin models to investigate the relationship between the dynamic behavior and equilibrium phase transitions.

\begin{acknowledgments}
This work was financially supported from the German Science Foundation (DPG) within the Graduiertenkolleg GRK 1885. We thank A.\ Peter Young for many interesting discussions and helpful suggestions.
\end{acknowledgments}

\bibliography{references}

\end{document}